\documentclass[journal]{IEEEtai}

\usepackage[colorlinks,urlcolor=blue,linkcolor=blue,citecolor=blue]{hyperref}

\usepackage{color,array}

\usepackage{graphicx}

\usepackage[latin5]{inputenc}
\usepackage{times}
\usepackage[figuresright]{rotating}
\usepackage{float}
\usepackage{subcaption}
\usepackage{url}
\usepackage{setspace}
\usepackage{balance}
\usepackage{multirow}
\usepackage{color}
\usepackage{epsfig}
\usepackage{amsmath, amssymb}

\usepackage{capt-of}
\usepackage{multirow}
\usepackage{array}
\usepackage{caption} 
\usepackage{pslatex}
\begin{document}

\title{Redefining Wireless Communication for 6G: \\Signal Processing Meets Deep Learning with Deep Unfolding}

\author{\IEEEauthorblockN{Anu Jagannath, \textit{Member}, IEEE, Jithin Jagannath, \textit{Senior Member}, IEEE, and Tommaso Melodia, \textit{Fellow}, IEEE}
\thanks{Anu Jagannath and Jithin Jagannath is with Marconi-Rosenblatt AI/ML Innovation Laboratory, ANDRO Computational Solutions, LLC, Rome, USA.} 
\thanks{Anu Jagannath is also with Institute for the Wireless Internet of Things, Northeastern University, Boston, USA. Jithin Jagannath is also affiliated with Department of Electrical Engineering at the University at Buffalo, State University of New York. (e-mail: \{ajagannath,jjagannath\}@androcs.com).}
\thanks{Tommaso Melodia is with the Institute for the Wireless Internet of Things, Northeastern University, Boston, USA (e-mail: melodia@northeastern.edu).}
}
\markboth{Journal of IEEE Transactions on Artificial Intelligence, Vol. 00, No. 0, Month 2020}
{First A. Author \MakeLowercase{\textit{et al.}}: Bare Demo of IEEEtai.cls for IEEE Journals of IEEE Transactions on Artificial Intelligence}

\maketitle

\begin{abstract}
The year 2019 witnessed the rollout of the 5G standard, which promises to offer significant data rate improvement over 4G. While 5G is still in its infancy, there has been an increased shift in the research community for communication technologies beyond 5G. The recent emergence of machine learning approaches for enhancing wireless communications and empowering them with much-desired intelligence holds immense potential for redefining wireless communication for 6G. The evolving communication systems will be bottlenecked in terms of latency, throughput, and reliability by the underlying signal processing at the physical layer. In this position paper, we motivate the need to redesign iterative signal processing algorithms by leveraging deep unfolding techniques to fulfill the physical layer requirements for 6G networks. To this end, we begin by presenting the service requirements and the key challenges posed by the envisioned 6G communication architecture. We outline the deficiencies of the traditional algorithmic principles and data-hungry deep learning (DL) approaches in the context of 6G networks. Specifically, deep unfolded signal processing is presented by sketching the interplay between domain knowledge and DL. The deep unfolded approaches reviewed in this article are positioned explicitly in the context of the requirements imposed by the next generation of cellular networks. Finally, this article motivates open research challenges to truly realize hardware-efficient edge intelligence for future 6G networks. 
\end{abstract}

\begin{IEEEImpStatement}
In this article, we discuss why the infusion of domain knowledge into machine learning frameworks holds the key to future embedded intelligent communication systems. Applying traditional signal processing and deep learning approaches independently entails significant computational and memory constraints. This becomes challenging in the context of future communication networks such as 6G with significant communication demands where dense deployments of embedded internet of things (IoT) devices are envisioned. Hence, we put forth deep unfolded approaches as the potential enabling technology for 6G Artificial Intelligence (AI) radio to mitigate the computational and memory demands as well as to fulfill the future 6G latency, reliability, and throughput requirements. To this end, we present a general deep unfolding methodology that can be applied to iterative signal processing algorithms. Thereafter, we survey some initial steps taken in this direction and more importantly discuss the potential it has in overcoming challenges in the context of 6G requirements. This article concludes by providing future research directions in this promising realm. 
\end{IEEEImpStatement}


\section{Introduction}\label{sec:introduction}
Fifth-generation (5G) mobile technology made significant advancements with respect to previous generations. The year 2019 witnessed the Phase-I 5G rollouts with a promise of fulfilling the needs of end-users and network operators. While 5G is still in the rollout and evaluation phase, current research activity is centered around beyond-5G communications to meet the ever-growing demands of tactile communication with immersive user experience. The ITU-Telecommunication standardization sector organized the Network 2030 focus group to study the capabilities of the network for the year 2030 and after to support revolutionary communication technologies. Network 2030 intends to identify the enabling technologies and infrastructure evolutions to offer revolutionary communication experience with immersive holographic communication, telesurgery, tactile multimedia communication over the internet, among others. The enabling technologies require high-speed communication with ultra-low latencies to guarantee tactile internet with an immersive communication experience. With this vision, the Future Communications Summit has been organized with the support of IEEE to organize research workshops and layout the R\&D roadmap for enabling technologies and services towards 6G. The race towards 6G has already begun with network operators, commercial cellular companies, and academic institutions leading the revolutionary beyond 5G research. For instance, 6G-Enabled Wireless Smart Society and Ecosystem - 6Genesis - was launched in 2018 by the Academy of Finland to study the 6G service requirements and the key technology enablers \cite{6Genesis}.

\begin{figure*}[t!]
\centering
\epsfig{file=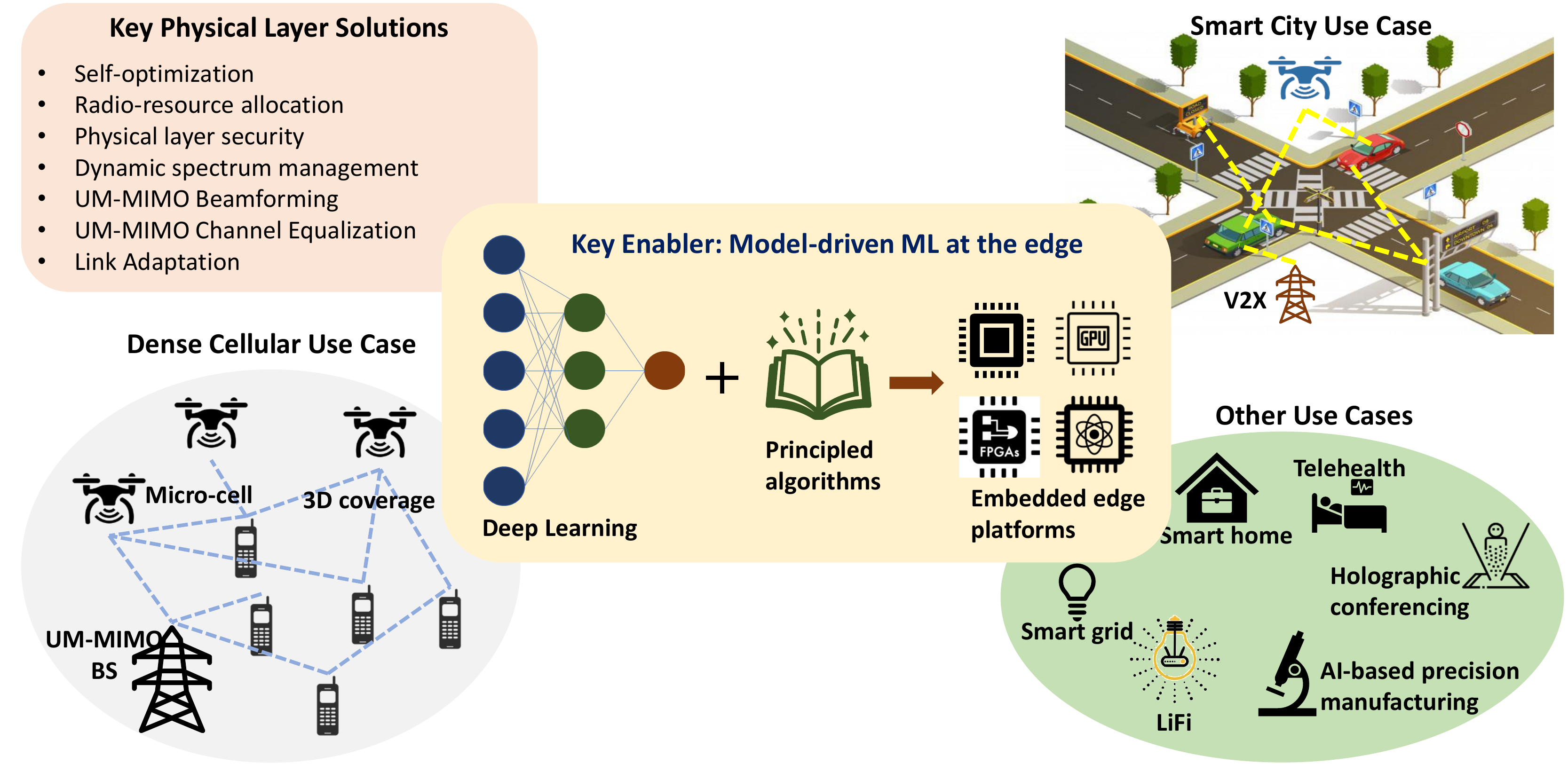, width=6.3 in,}
\caption{Hardware-efficient ML at the edge}\label{fig:6G}
\end{figure*}

\begin{figure}[t!]
\centering
\epsfig{file=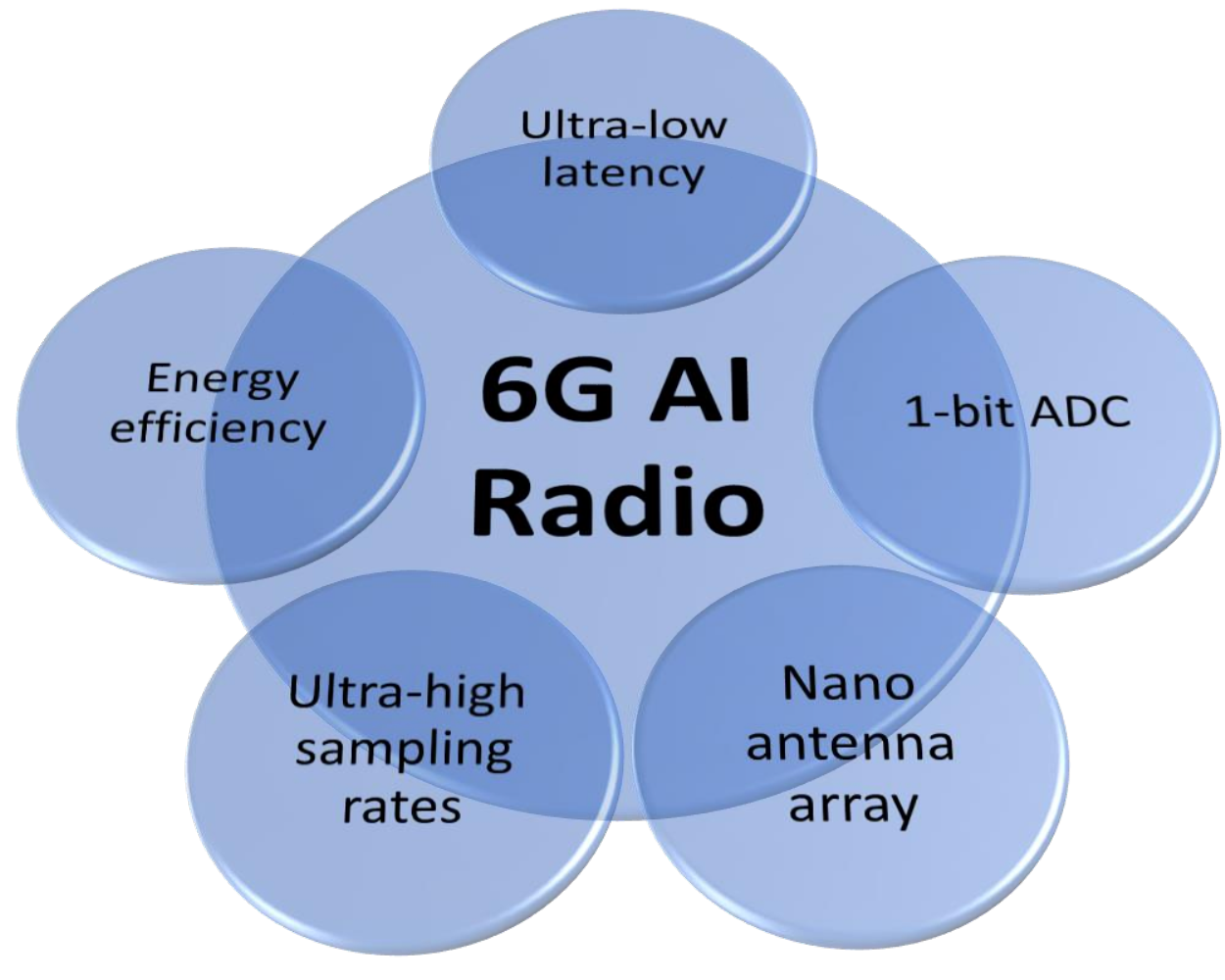, width=2.4 in,}
\caption{Overview of hardware requirements of 6G AI radio}\label{fig:airadio}
\end{figure}

The beyond 5G research will encompass diverse domains including scalable-intelligent communications, autonomous network operations, intelligent surfaces, physical layer approaches, etc. The peak data rate of 6G is envisioned to be 1 Tbps which is a $100\times$ improvement over 5G. The diverse application scenarios of 6G as in Fig.\ref{fig:6G} will require satisfying heterogeneous service requirements \cite{6Gcomms,6Gfrontier,6Greq,6Gwhat,6gsaad} for machine-to-machine, human-to-machine, ultra-dense IoT networks, etc., as shown in Table \ref{tab:tabl}. Figure \ref{fig:6G} shows a bird's eye view of how in each envisioned 6G use case deep unfolding physical layer approach can serve as the key enabler. The envisioned 6G communication will witness 3D communication hierarchy, dense device deployments to support multiple revolutionary autonomous use cases such as holographic video conferencing applications, connected autonomous scenarios like smart grid, smart city, precision manufacturing, vehicle-to-everything (V2X), telehealth, etc. Consequently, the envisioned hardware requirements of the future 6G AI radio is shown in Fig.\ref{fig:airadio}.

\begin{table}[!h]
\caption{Emerging communication requirements for 6G}
\centering
\def\arraystretch{1.3}%
\begin{tabular}{|p{3.0 cm}|p{4.3 cm}|}
\hline
\textbf{Specifications}       & \textbf{Requirements}      \\ \hline
Ultra-high data rate        & 1 Tbps (downlink) \newline 100 Gbps (user data rate)          \\ \hline
Ultra-low latency        & $\sim$ 0.1 ms            \\ \hline
Processing delay & $\leq$ 10 ns \\ \hline
Ultra dense network        & $10\times$ w.r.t 5G device/km$^2$               \\ \hline
Mobility & 1000 km/hr \\ \hline
Reliability &$10^{-9}$ Frame error rate \\ \hline
Spectral efficiency &100 bps/Hz (downlink)\\ \hline
Energy efficiency        & $\sim1$pJ/bit \\ \hline
\end{tabular}
\label{tab:tabl}
\end{table}  

Intelligent communication is envisioned as a key enabler for future 6G networks \cite{6Gvision}. Machine Learning (ML) has attained significant breakthroughs in domains such as natural language processing, speech recognition, computer vision, among others, and is emerging as a prominent tool for wireless communications. Deep learning (DL) \cite{Goodfellow-et-al-2016} - a subset of ML - has been applied to wireless communication problems such as signal recognition \cite{Jagannath19MLBook,DLoshea, AJagannath21ICC}, detection, characterization, channel estimation, optimal network resource allocation \cite{JagannathAdHoc2019}, error correction coding schemes, and other physical layer applications \cite{ML30,mlair,RestucciaPolymorf}. In addition to physical layer optimizations, DL has been applied to upper layers as well for intelligent routing, MAC, and transport control. Although from a 6G latency and data speed requirements perspective, the upper layer enhancements would be constrained by the physical layer signal processing capability. 

Several works \cite{inteledge,roadmap,ML30,mlair} have studied the concept of intelligent wireless communication. An intelligent edge concept for wireless communication is elaborated in \cite{inteledge}. The learning-driven radio resource management and signal encoding problem is studied in depth. However, they do not study the complexity of these approaches and their implementation platforms (edge computing devices). The architecture of 6G networks and the concept of intelligent radio is presented in \cite{roadmap}. They also briefly mention the importance of an intelligent physical layer in realizing the 6G communication requirements. Their work does not present the candidate physical layer approaches that will realize the intelligent physical layer concept. A detailed study of the traditional machine learning techniques as applied to solving wireless communication problems is discussed in \cite{ML30}. Nonetheless, the performance of these techniques from a hardware-efficient implementation standpoint is lacking in this work. Authors of \cite{mlair} focus on various machine learning approaches for physical layer communication but do not address model driven deep unfolding which is the focus of this article in the context of envisioned 6G requirements. 

Previous works especially \cite{6Gcomms,6Gvision,6G_Giordani,6Gspec} discussed the various 6G use cases and enabling technologies. However, they do not mention how physical layer techniques can be realized for an edge computing platform to facilitate the key concept of ML at the edge for 6G. Recently, \cite{unfoldBalat} surveyed deep unfolding as applied to physical layer signal processing. The focus of this work was to provide an in-depth survey of deep unfolding techniques with special emphasis on multiple-input multiple-output (MIMO) wireless systems and belief propagation decoding of error correction codes. In contrast, our work provides a succinct account of the state-of-the-art deep unfolding approaches which could serve as potential 6G enablers, contrast, and tabulate their performance with traditional principled approaches as well as other deep unfolded counterparts. We present the deep unfolded techniques that perform various receiver operations such as signal estimation and detection, self-interference cancellation, and advanced error correction. In addition, we present the generic layout of deep unfolding methodology for any signal processing application and discuss revolutionary research directions to enable future communication networks. 

In summary, we will be focusing on the promising physical layer solutions that can open doors for future 6G networks. The data-hungry nature and exponential complexity of DL solutions are the key challenges hindering its widespread deployment. 
We aim to motivate the reader in understanding how computationally intensive DL approaches may not be the solution to achieve the 6G communication requirements. Instead, integrating the powerful learning capability of DL with algorithmic knowledge will relieve the computational burden as well as improve the robustness. Therefore, this is the first work that studies the potential intelligent hardware-efficient physical layer solutions that would serve as key enablers to realize the 6G AI radio concept (Fig.\ref{fig:airadio}).

\section{Background - 6G AI Radio: Need for Edge Intelligence}
We argue that AI will empower 6G in all aspects from network orchestration and management, physical layer signal processing, and data mining, to service-based context-aware communication. Researchers have already started envisioning and planning the key enabling technologies to support future 6G communications under different labels such as 5G+ and beyond 5G. uRLLC was defined in 5G standards to attain latency of the order of 1 ms for latency-critical applications. Fog networking and mobile edge computing paradigms were introduced in 5G to greatly 
reduce the delay as well as network congestion typically encountered in a user equipment (UE) to centralized base station communication. 
However, fog nodes cannot act as independent cloud data centers and rely on a centralized cloud. Cloud radio access networks (C-RAN) are another derivative of cloud computing wherein the traditional base stations are replaced with distributed remote-radio-heads and a centralized baseband unit (BBU). In the C-RAN architecture, the signal processing computations are performed at the BBU. 
Although C-RAN may fulfill the service requirements of 5G, the communication overhead, service heterogeneity, and computations performed at the BBU to provision 6G networks will prohibitively increase the computational complexity and latency. 
An emerging idea to address this is the open radio access network (O-RAN) which embraces openness and intelligence as its core concepts \cite{oran}. O-RAN provisions well-defined open interfaces between elements implemented on general purpose hardware as well as the integration of RRHs and BBUs from different vendors. We envision model-driven intelligent signal processing modules deployed on the O-RAN for inference with computationally simpler hardware. 


Cell densification is a key enabler for attaining increased network capacity. 6G aims at ultra-dense network deployments (ultra-massive IoT) involving multiple cells within a macro-cell. The diverse 6G use cases such as high precision manufacturing, vehicle-to-everything, smart homes will involve heterogeneous devices operating in a micro-cell. However, such 6G applications will involve addressing several key challenges. The heterogeneity, high-frequency operation, mobility, and dense operation will introduce a different dimension of propagation, spectrum access, radio resource allocation, scheduling, and security concerns. These hurdles will be exacerbated with the envisioned 3D communication infrastructure \cite{6G_Giordani} for 6G incorporating mobile ground and aerial platforms. In the large scale dense deployments planned for 6G, intelligent physical layer schemes will play a crucial role in fulfilling the service requirements. 

The exploitation of ML at the edge will become a primary enabler for 6G communications. State-of-the-art DL solutions for wireless communication including multiple-input multiple-output (MIMO) channel estimation and equalization, beamforming, error correction and coding, signal recognition, etc., \cite{JagannathAdHoc2019} are computationally complex requiring powerful computational platforms. However, the adept learning capability of DL architectures motivates the need to incorporate them for future communication networks. In fact, significant breakthroughs in AI have urged it to be part of 5G yet deployed only in facilities with massive training data and powerful computing platforms. However, to facilitate ML at the edge, lighter implementations of DL solutions will need to be developed. Additionally, the black-box nature of neural networks (NNs) renders them incomprehensible and unpredictable such that it is challenging to gain insight into the learned function from the network architecture. Therefore, NNs introduce three key challenges: \emph{resource-constraint}, \emph{computational complexity}, and \emph{black-box nature}.
The baseband physical layer signal processing techniques will be run on the UE as well as base stations. Every generation of communication standards is bottlenecked by the UE's capability. Consequently, executing these DL solutions in the current UE platforms will be infeasible as they do not possess powerful computational platforms as required by these approaches. Therefore, intelligent baseband signal processing designed for future 6G networks must be tailored for lightweight embedded computational platforms with stringent energy efficiency, reliability, and latency objectives. 
Fusing domain knowledge into the DL architectures holds immense potential to accelerate the training process and model convergence while enhancing the model efficiency. To this end, \emph{deep unfolding} \cite{Hershey,unfoldBalat} - algorithmic unrolling of signal processing models (principled approaches) integrated with trainability using DL strategies - is emerging as a scalable and efficient approach. We envision such deep unfolded (model-driven) approaches 
as a promising solution in facilitating ML at the edge for future 6G end-to-end communication architectures to garner performance gains and implementation ease. Model-driven DL will form a pervasive component of the 6G communication infrastructure granting the ability to perform self-optimization of network resources, traffic control, authentication measures, anomaly prediction and preemption, rapid channel estimation, error correction and coding, and physical layer security schemes forming a fully \emph{autonomous} 6G communication grid.
\section{Methodology - Deep unfolded signal processing for 6G}
\label{sec:du}

\begin{figure}[t!]
\centering
\epsfig{file=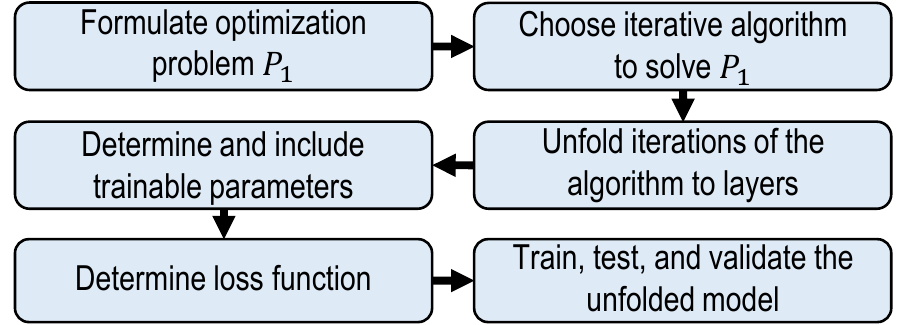, width=3.4 in,}
\caption{Methodology for Deep unfolded signal processing}\label{fig:steps}
\end{figure} 

\begin{figure*}[h!]
\centering
\epsfig{file=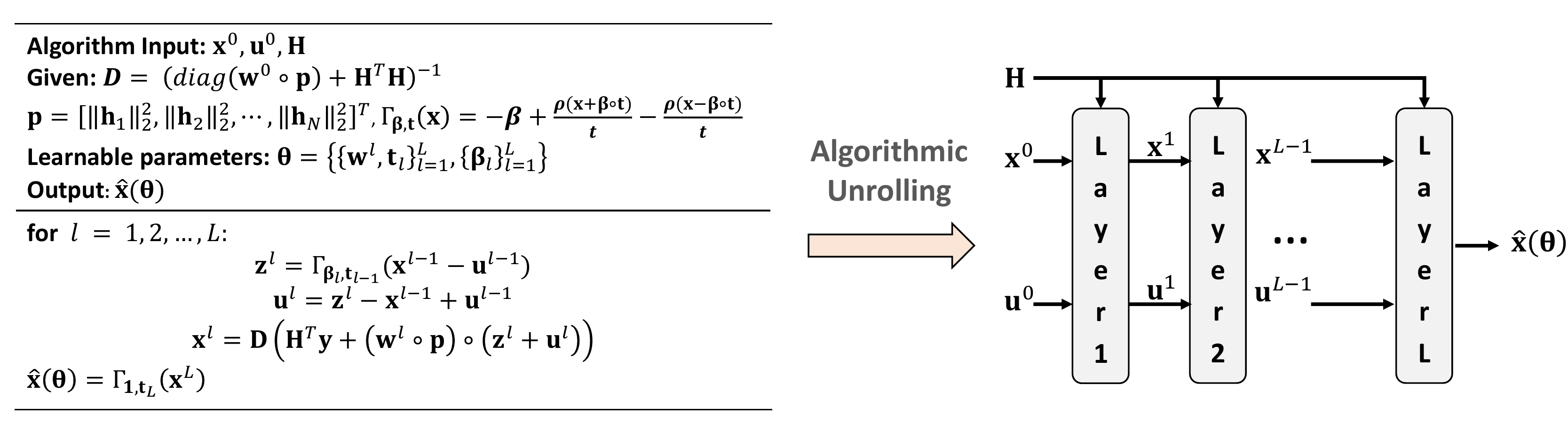, width=7 in,}
\caption{Deep unfolded representation of ADMM-Net}\label{fig:admm-net}
\end{figure*}

We believe model-driven DL could be the key to enhance the performance and inference times of communication systems. This section discusses the advantages of combining domain knowledge with the learning ability of DL to mitigate the deficiencies of traditional signal processing and black-box NN approaches.

More than a handful of principled signal processing approaches at the physical layer - signal detection, interference estimation, channel estimation, encoding, decoding, etc - can be defined by optimization problems which can be subsequently solved with iterative inference algorithms. Such iterative approaches typically involve computationally intensive operations such as eigen decomposition, matrix inversion, etc., requiring a large number of iterations to converge. One such example of a non-scalable traditional approach is maximum-likelihood detection which attains optimal performance but at the expense of exponential complexity that scales with the decision variables. However, the computationally lighter alternative suboptimal linear detectors such as linear minimum mean squared error (MMSE) and Zero-Forcing comes with reduced reliability. Similarly, the traditional iterative MIMO detectors such as Approximate Message Passing (AMP) and Expectation Propagation offer good reliability but with moderate computational complexity.

Traditional signal processing approaches rely on careful parameter tuning, initialization, and step-size selection to offer acceptable performance and convergence speed. In practice, they are tuned based on heuristics such that they are chosen arbitrarily or from exhaustive searches in simulations. However, such heuristic-based selections result in instability and suboptimal performance. We argue that such deficiencies can be mitigated by combining the domain knowledge from principled signal processing algorithms with the learning ability of NNs to yield \emph{deep unfolded signal processing}. This can be perceived as an instance of \emph{model-driven NN}.


Deep unfolding refers to the process of unfolding the iterations of a principled inference algorithm to form a layered structure analogous to NN. Deep unfolded signal processing combines the benefits of both DL and domain knowledge of the signal processing models to improve the model performance with computationally lighter architectures. For example, an $N$-step iterative inference algorithm can be unfolded into an $N$-layered NN with trainable parameters based on the model. The parameters can be learned with tools from DL such as backpropagation, SGD, etc. A general layout to perform deep unfolding 
is shown in Fig. \ref{fig:steps}.

As an illustrative example, we will show the algorithmic unrolling of the Alternating Direction Method of Multipliers (ADMM) as proposed in \cite{ADMM_MIMO} for MIMO detection in Fig. \ref{fig:admm-net}. Here, the optimization problem $P_1$ is ADMM. The $L$-step iterative algorithm, unfolding, and trainable parameters are shown in Fig. \ref{fig:admm-net}. ADMM-Net considers a MIMO system with received signal $\mathbf{y}\in \mathbb{R}^M$, transmit signal $\mathbf{x}\in \{\pm1\}^N$, and channel matrix $\mathbf{H}\in \mathbb{R}^{M\times N}$. A weighted mean squared error between ground truth and predicted symbols is chosen as the loss function to train ADMM-Net. Here, $\Gamma_{\mathbf{\beta,t}}(\mathbf{x})$ is the projection operator. All these components together help to successfully unroll the iterative ADMM algorithm into a neural network architecture shown in Fig. \ref{fig:admm-net}. The application and advantages of ADMM-Net are discussed in greater depth in section \ref{sec:mimo}.

\section{Discussion}\label{sec:disc}

In this section, we will review physical layer pertinent signal processing tasks that have been significantly improved with deep unfolding. We will specifically discuss this in the context of envisioned 6G requirements and how some of these steps will be critical for the next generation of communication networks. 

\subsection{Signal Recovery Schemes}
Signal recovery involves the problem of reconstructing the signal from noisy measurements. This could involve suppressing the effect of self-interference, co-channel interference, noise, multipath, or propagation effects from the received signal. Several key techniques envisioned for 6G such as mmWave massive MIMO, Terahertz (THz) band communication, optical wireless communication, full-duplex (FD), ultra-massive MIMO (UM-MIMO), etc., require channel estimation and interference suppression techniques to sustain reliable communication links. Such potential signal reconstruction techniques for 6G must be adaptive, fast, and reliable to sustain the latency and data rate requirements. Consequently, we investigate current state-of-the-art deep unfolded signal recovery techniques that could serve as potential candidates or form a stepping-stone for future enhancements. Deep unfolding has been applied to signal recovery problems recently in \cite{TISTA,deeprec,Kurzo2018_DUIntCancel}. Sparse signal processing will be required to support tactile internet applications for 6G with latency $\sim$0.1ms. Massive sporadic traffic generated from dense IoT devices requires ultra-fast burst data acquisition and processing at the physical layer.
Iterative Soft Thresholding Algorithm (ISTA) is a powerful signal processing tool for sparse-signal recovery. However, ISTA requires numerous iterations to converge at an acceptable normalized mean squared error (NMSE) which can lead to processing delay contributing to communication latency. Several successors of ISTA - learned ISTA (LISTA), AMP, Trainable Iterative Soft Thresholding Algorithm (TISTA) - were proposed recently of which the latter have been shown to outperform the others \cite{TISTA}. 

TISTA deep unfolded the iterative ISTA with a trainable step-size parameter. Additionally, the thresholding function of ISTA is replaced with an MMSE-based shrinkage function. TISTA essentially unfolded the $N$-step ISTA into an $N$-layered DL architecture with $N+2$ learnable parameters. 
The fewer trainable parameters lead to a highly stable and faster training process. TISTA adopts an incremental training strategy to mitigate the vanishing gradient problem. TISTA demonstrated significantly faster convergence than orthogonal approximate message passing (OAMP) and LISTA \cite{LISTA}. Specifically, TISTA exhibited 37$\times$ faster performance and better NMSE than LISTA. 
The computational efficiency was demonstrated by 
evaluating TISTA on an Intel Xeon(R) 6-core CPU rather than GPU. 

Energy efficiency will be a primary factor in designing future 6G radios. Analog-to-digital converter (ADC) is a ubiquitous component in radio hardware. However, their energy consumption and chip area increase with their bit resolution. Especially, to support the ultra-high data rate 6G communication technologies such as THz and optical wireless communication, ADCs that can support a very high sampling rate will become quintessential. Hence, even an 8-bit ADC to support such high sampling rates will significantly scale the manufacturing cost and power consumption of the device. However, very low-resolution 1-bit ADCs that can support very high sampling rates can significantly lower the power consumption, cost as well as the chip area. Consequently, signal processing techniques for 1-bit quantized signals must be considered for future transceiver architectures. An unfolded deep NN-based signal recovery scheme for 1-bit quantized signals - DeepRec - is proposed in \cite{deeprec}. DeepRec unfolds the iterations of the maximum-likelihood signal estimator into the layers of a deep NN. The maximum-likelihood estimation can be solved with the iterative gradient ascent method. Each iteration of the gradient ascent was represented with an equivalent layer in NN with ReLU activation. For a 90-layer DeepRec model, the performance improvements in terms of NMSE and computational efficiency were demonstrated in contrast to the traditional gradient descent method.

Promising technologies such as THz in conjunction with FD radios have the potential to fulfill the Tbps data rate requirements of 6G. Radio transceivers with the ability to transmit and receive simultaneously on the same frequency band - FD radios - have the potential to \emph{double the spectral efficiency} of a point-to-point radio link. Consequently, FD radios have the potential to double the attainable data rates of future communication networks. 
However, the self-interference (SI) caused by such FD radios is a serious limiting factor that is slowing down its widespread adoption.

SI mitigation involves estimating the interference term and suppressing it from the signal component. The performance gains from deep unfolding a state-of-the-art polynomial SI cancellation approach are investigated in \cite{Kurzo2018_DUIntCancel}. The weighted polynomial sum expression is unrolled into a feed-forward NN with one hidden layer. 
Deep unfolding yielded a computationally simpler architecture with only 13 neurons (nodes) whose weights can be estimated by supervised learning with backpropagation.
In contrast to the polynomial canceller, the unfolded SI canceller \cite{Kurzo2018_DUIntCancel} exhibited a lower quantization bit-width requirement for the same cancellation performance. The computational efficiency and hardware implementation were demonstrated on Xilinx Virtex-7 FPGA and Fully-Depleted-Silicon-On-Insulator (FDSOI) ASIC. The FPGA implementation demonstrated a 96\% higher throughput and significantly lower resource utilization in contrast to polynomial canceller. Similarly, the unfolded SI canceller implementation on ASIC exhibited an 81\% better hardware efficiency. 

\subsection{MIMO Detection Techniques}\label{sec:mimo}
Massive MIMO techniques in the mmWave band are serving as key enablers of the 5G networks. However, UM-MIMO techniques in conjunction with high frequency communication bands will serve as candidate technologies to satisfy the ultra-high data rate requirements of 6G communications. Large intelligent surfaces that scale beyond conventional antenna arrays are also envisioned as a candidate technology for 6G to enable wireless charging capabilities and extremely high data rates. Such intelligent surfaces can be realized via THz UM-MIMO. Fast, robust, and hardware-efficient MIMO transceiver techniques will be an inevitable part of future 6G systems. Therefore, we will review intelligent deep unfolded MIMO transceiver techniques that outperform traditional schemes.
Traditional MIMO detection techniques such as AMP, lattice decoding, sphere decoding, conditional maximum-likelihood decoding, etc., will incur significant complexity-reliability trade-off for large scale massive MIMO systems. For instance, the sphere decoder is a search algorithm that performs maximum-likelihood detection but with exponential complexity that scales with the number of transmit antennas. However, such exponential complexity will be detrimental to the latency and hardware efficiency of UM-MIMO systems envisioned for 6G. 
ADMM is an iterative algorithm to solve the maximum-likelihood MIMO detection problem which becomes computationally complex for large-scale MIMO systems.

ADMM-Net proposed in \cite{ADMM_MIMO} unfolds the iterative ADMM algorithm into a simpler DNN architecture. ADMM-Net is unfolded into a neural network with 40 layers to perform signal detection for a $160\times160$ massive MIMO system. The unfolding is performed by untying the penalty parameter $\mathbf{\lambda}$ into two terms such that one accounts for channel gain ($\mathbf{p}$) while the other serves as the trainable parameter ($\mathbf{w}$), i.e., $\mathbf{\lambda}=\mathbf{p}\circ\mathbf{w}$. The projection ($\Pi_{{\{\pm1\}}^{N}}$) at each layer ($l$) is untied to act as a per-layer trainable parameter ($\Gamma_{\mathbf{\beta}_l,\mathbf{t}_{l-1}}(\mathbf{x})$) as follows.
\begin{align}
 \mathbf{z}_l &= \Pi_{{\{\pm1\}}^{N}}\Big( \mathbf{x}_{l-1} - \mathbf{u}_{l-1}\Big)\; \text{in iterative ADMM}\\
    \mathbf{z}_l &= \Gamma_{\mathbf{\beta}_l,\mathbf{t}_{l-1}}\Big( \mathbf{x}_{l-1} - \mathbf{u}_{l-1}\Big)\; \text{in ADMM-Net}
\end{align}
The approach only involves a matrix inversion once initially instead of at each layer to ease the computational burden.
ADMM-Net was deployed on an Intel core i7-CPU to exhibit support on a computationally lighter platform with an inference time of $\sim 2.8$ ms whereas the semi-definite relaxation (SDR) and DetNet required 17.2 ms and 246 ms respectively for the $160\times160$ massive MIMO setting. Here, the traditional sphere decoder system was too slow to run to completion. Further, ADMM-Net outperformed traditional zero-forcing detector, SDR, and DetNet in terms of reliability for the $160\times160$ massive MIMO setting.

The past couple of years witnessed an emergence of several deep unfolded massive MIMO detection algorithms - DetNet, OAMPNet \cite{oampnet}, MMNet \cite{mmnet}, etc.,- that integrate the benefit of domain knowledge and DL. Among which MMNet outperforms its deep unfolded counterparts (DetNet, OAMPNet) as well as traditional model-based algorithms such as MMSE, Vertical-Bell-Laboratories-Layered-Space-Time (V-BLAST), SDR, and AMP under realistic MIMO channel conditions. MMNet models the iterative procedure to solve maximum-likelihood estimation as two separate NN architectures for simple AWGN and arbitrary channel matrices. To further illustrate this, consider a $N_t\times N_r$ MIMO system $\mathbf{y}=\mathbf{H}\mathbf{x}+\mathbf{n}$ with $N_t$ transmit antennas and $N_r$ receive antennas, where $\mathbf{y}\in\mathbb{C}^{N_r},\; \mathbf{x}\in\mathbb{C}^{N_t},\; \mathbf{H}\in\mathbb{C}^{N_r\times N_t}$, and $\mathbf{n}\sim \mathcal{CN}(0,\sigma^2\mathbf{I}_{N_r})$ are the received signal, transmitted signal, channel matrix, and additive white Gaussian noise respectively. The maximum-likelihood of $\mathbf{x}$ is $\hat{\mathbf{x}} = \arg\underset{\mathbf{x}\in\mathbb{C}^{N_t}}\min ||\mathbf{y}-\mathbf{Hx} ||_2$. A general iterative framework to solve the maximum-likelihood MIMO detection comprise,
\begin{align}
    \mathbf{z}_l &= \hat{\mathbf{x}}_l + \mathbf{A}_l(\mathbf{y}-\mathbf{H}\hat{\mathbf{x}}_l+\mathbf{b}_l)\; &\text{intermediate signal}\\
    \mathbf{x}_{l+1}&=\eta_l(\mathbf{z}_l)\;&\text{denoiser}
\end{align}
The MMNet neural network for arbitrary channel matrices corresponding to the above shown iterative framework is
\begin{align}
    \mathbf{z}_l &= \hat{\mathbf{x}}_l + \mathbf{\theta}_l^{(1)} (\mathbf{y}-\mathbf{H}_l\hat{\mathbf{x}}_l)\\
    \mathbf{x}_{l+1}&=\eta_l(\mathbf{z}_l;\mathbf{\sigma}_l^2)
\end{align}
where $\mathbf{\theta}_l^{(1)}$ is a $N_t\times N_r$ complex-valued trainable matrix and $\mathbf{\sigma}_l^2 = f(\mathbf{\theta}_l^{(2)})$ represents the noise variance with a trainable vector $\mathbf{\theta}_l^{(2)}$ of size $N_t\times 1$. In its simplistic form, each layer of the network performs two steps: 1. obtain an intermediate signal representation using a signal estimate from the previous layer, residual term, and bias. 2. Apply a non-linear denoising function on the intermediate signal to obtain signal estimate which will be fed as input to the subsequent layer. 

A 10-layer MMNet model outperforms the classic MMSE detector and OAMPNet by $\sim 4-8$dB and 2.5dB respectively. Additionally, MMNet achieves the SNR gain at 10-15$\times$ lower complexity in contrast to OAMPNet. The traditional SDR and V-BLAST schemes deviate from the ideal maximum-likelihood performance at higher SNR and modulation schemes. Similarly, the AMP algorithm suffers from robustness issues at higher SNR levels and modulation orders. However, MMNet with its lower complexity and fewer trainable parameters stays very close (within 1.5dB) to the maximum-likelihood performance with increasing SNR and modulation orders.  

\begin{figure}[t!]
\centering
\epsfig{file=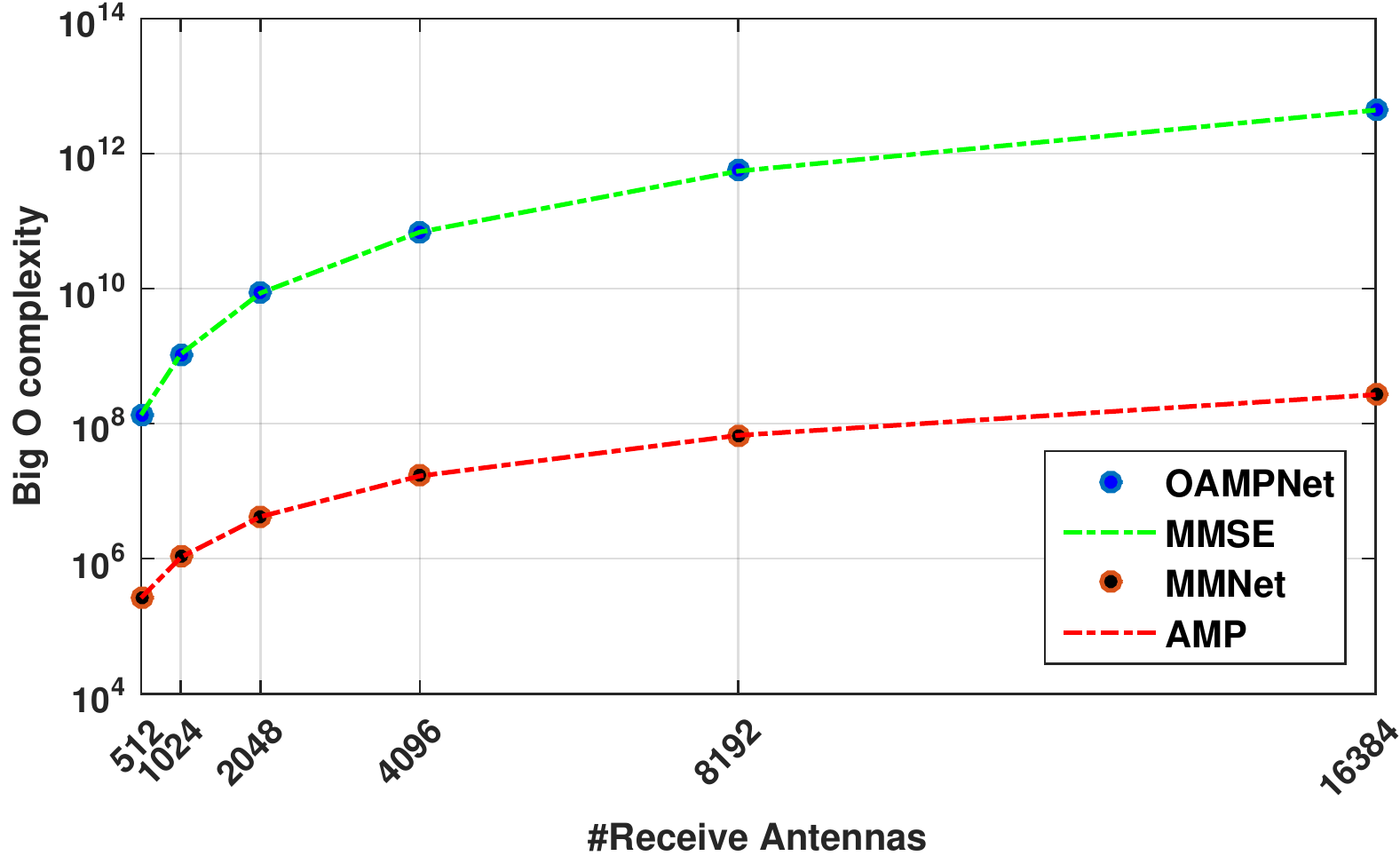, width=3.2 in,}
\caption{Complexity comparison for 6G UM-MIMO scenarios}\label{fig:bigo}
\end{figure}

The space dimension will be significantly exploited in 6G networks in conjunction with high-frequency bands. For example, a UM-MIMO array for THz communications in the 1 THz band will involve at least $1024\times1024$ antenna elements. Figure \ref{fig:bigo} shows the Big-O complexity analysis of traditional AMP and MMSE with two deep unfolded techniques - MMNet and OAMPNet for varying UM-MIMO antenna array configurations. MMNet offers lower complexity as compared to MMSE and OAMPNet while offering lightweight complexity as with traditional AMP in addition to performance enhancements as discussed previously. Intelligent deep unfolding approaches adopted in these MIMO receiver techniques lay the basis for other multi-antenna techniques such as beamforming, spatial multiplexing, and space-time block coding. We expect such unique combinations of intelligent multi-antenna techniques when integrated with high-frequency bands to profoundly benefit the 6G communication networks in attaining the desired range as well as communication capacity.
\subsection{Advanced Error Correction Schemes}
Resilient communication is paramount in wireless networks. Especially, with the envisioned dense networks, propagation and multipath channel effects associated with high frequency bands, error detection and correction schemes become inevitable to sustain the future 6G wireless links. With this vision, European Union founded a 6G research project - Hexa-X - for advanced error correction, distributed MIMO, applied AI/ML, channel coding, adaptive modulation schemes to enable basic 6G technologies \cite{hexa}. The flagship initiative headed by Nokia and Ericsson brings together a consortium of major industry and academic stakeholders to direct research efforts to standardize 6G. Along the same line, another EU project Horizon2020 ICT-09-2017 ``Networking research beyond 5G'' aimed at studying key enabling technologies such as error correction schemes, THz communication, etc for future communication networks.
\begin{table*}[!ht]
\caption{Review of promising direction towards ML at the edge}
\centering
\small
\def\arraystretch{1.3}%
\begin{tabular}{|p{2.6 cm}|p{3.8cm}|p{3.5cm}|p{5.5cm}|}
\hline
\textbf{Deep unfolded\newline approaches}       & \textbf{Benchmark approaches}  & \textbf{Implementation Platform}  & \textbf{Performance metrics}      \\ \hline
TISTA\cite{TISTA}        & LISTA, AMP, OAMP & Intel Xeon(R) 6-core CPU &Convergence rate           \\ \hline
DeepRec\cite{deeprec}        & Gradient descent & Quadcore 2.3 GHz CPU &Reliability, computational efficiency            \\ \hline
Unfolded \newline SI-canceller\cite{Kurzo2018_DUIntCancel} &Polynomial canceller &Xilinx Virtex-7 FPGA, \newline FDSOI ASIC        &Quantization bitwidth, hardware efficiency,\newline hardware resource utilization \\ \hline
ADMM-Net\cite{ADMM_MIMO} &Sphere decoder, Zero-forcing detector, SDR, DetNet &Intel Core i7-5820K CPU & Runtime, reliability \\ \hline
MMNet\cite{mmnet} &MMSE, V-BLAST, SDR, AMP, DetNet, OAMPNet &GPU &Complexity, reliability\\ \hline
TPG decoder\cite{ldpc} &Belief propagation &Not specified        &Reliability, SNR gain, complexity \\ \hline
TurboNet\cite{turbonet}        & MAP, max-log-MAP, \newline NN-based BCJR &NVIDIA GeForce GTX 100 Ti GPU &Complexity, reliability \\ \hline
\end{tabular}
\label{tab:tabl2}
\end{table*}  

Low-density parity check (LDPC) codes introduced by Gallager has been adopted in 5G networks especially in the eMBB scenarios. 
However, to support the 100$\times$ throughput requirements of 6G, the encoder and decoder for such advanced error correction schemes must further improve in terms of latency and hardware efficiency. A model-driven DL approach to LDPC decoding intended to support NN oriented AI chips is proposed in trainable projected gradient descent (TPG decoder) \cite{ldpc}. 
A linear programming (LP) formulation of LDPC decoder was proposed by J. Feldman which can be represented as an optimization problem. The LP optimization problem can be reformulated to an unconstrained setting by including a penalty function ($P(\mathbf{x})$) in the objective as $f_{\beta}(\mathbf{x})=\mathbf{\lambda x}^T + \beta P(\mathbf{x})
$, where $\mathbf{x} = 1-2\mathbf{c}$ is the bipolar codeword sent over the channel, $\mathbf{c}$ is the binary codeword, and $\lambda$ is the log-likelihood ratio vector of received codeword $\mathbf{y}$.
This reformulated unconstrained optimization problem can be solved with the well known projected gradient descent (PGD) algorithm. PGD comprises of two steps - gradient and projection, shown below,
\begin{align}
    \mathbf{r}_l &= \mathbf{s}_l-\gamma_lf_{\beta_l}(\mathbf{s}_l) \;\text{gradient step}\\
    \mathbf{s}_{l+1} &= \epsilon(\alpha(\mathbf{r}_l-0.5)) \; \text{projection step}
\end{align}
where $\epsilon(\cdot)$ is the sigmoid function and $\alpha$ controls the softness of the projection. TPG decoder unfolds the iterations of the gradient step of the PGD algorithm. The projection is achieved with a non-linear activation function - sigmoid. The parameters in the gradient and projection steps are chosen as trainable parameters for the learning process to control the stepsize and penalty in the gradient step as well as the projection softness. The final parity check in the decoding is a thresholding function which yields the estimated codeword as $\hat{\mathbf{c}} = \theta(\mathbf{s}_{l+1})$. TPG decoder for a (3,6) regular LDPC was shown to outperform belief propagation with a 0.5dB gain at a BER=$10^{-5}$.

Turbo codes are among the advanced channel coding schemes that are employed in deep space communications, 3G/4G mobile communication in UMTS, and LTE standards. However, to exploit Turbo codes for 6G it will need to support significantly higher data rates at lower implementation complexity. Previously proposed recurrent NN-based BCJR algorithm relies on a large amount of training data and is computationally complex. 
Incorporating domain knowledge will ease the requirement of large training data as well as can potentially minimize the number of trainable parameters along with improved performance. TurboNet introduced in \cite{turbonet} combines algorithmic knowledge from traditional max-log maximum \emph{a posteriori} (MAP) with deep NN. 
Each iteration of max-log-MAP algorithm is represented by a deep NN-based decoding unit. The log-likelihood-ratio output of the final decoding unit is subject to a non-linear activation function (sigmoid) to constraint the output in the range of [0,1]. The soft bit decisions are hard-decision thresholded to obtain binary values. TurboNet with 3 decoding units for Turbo (40,132) and (40,92) codes was shown to outperform traditional MAP and max-log-MAP algorithms. TurboNet has significantly fewer parameters (17.8K) and faster computation in contrast to 3.85M parameters of NN-based BCJR. Additionally, TurboNet with only 3 decoding units exhibits lower latency in comparison to traditional max-log-MAP with 5 iterations.


To summarize, Table \ref{tab:tabl2} shows the model-driven approaches discussed in this section along with their benchmarked approaches and performance metrics. Here, we showed a glimpse of the current state-of-the-art model-driven signal processing approaches that find a balance between principled and DL approaches in terms of reliability and complexity. Such deep unfolded approaches leverage the synergy between domain knowledge and DL to deliver unprecedented capacity. Imparting domain knowledge offers predictable and interpretable performance to model-driven approaches. Additionally, fewer training parameters allow for faster training and convergence. We envision such complexity reduction and reliable performance to find prominence in realizing the ML at the edge aspect of the 6G communication infrastructure.

\section{Conclusion and Future Research Directions}
This article discussed the state-of-the-art deep unfolded signal processing approaches and motivated such techniques to realize 
ML at the edge for 6G communication systems. Computational simplicity, scalability, accelerated convergence, small memory footprint, and high reliability are key to realize the 6G communication objectives. 
TISTA \cite{TISTA} involved requiring prior distribution of the original signal. MMNet \cite{mmnet} considered the only source of randomness in noise, original signal, and channel during the study. TurboNet \cite{turbonet} and TPG decoder \cite{ldpc} were evaluated over AWGN channel. For larger LDPC codes, the number of search steps can have a significant impact on the receiver processing delay. Setting the number of search steps as a learnable parameter rather than fixing them as arbitrary can be an improvement over the TPG decoder. The convergence rate and computational complexity of the discussed approaches in presence of highly dynamic channel, mobility, and data traffic conditions as in a 6G communication network will need to be investigated further. Despite recent studies in deep unfolded signal processing approaches, there are still open problems that can be investigated in the future.

\textbf{Rapid online learning.} The 6G communication infrastructure opens a pandora's box of wireless communication challenges such as heterogeneous service requirements, dense device deployments, 3D communication architecture, etc. 
The physical layer approaches must therefore be rapid and adaptive to learn instantaneous and never-before-seen scenarios. Few-shot learning and meta-learning are newly christened paradigms in ML that enable exploiting prior knowledge in allowing a model to learn from a few scenarios. A promising direction towards rapid learning architectures for 6G would be to integrate domain knowledge with few-shot/meta-learning schemes to result in model-driven few-shot/meta-learning signal processing approaches.

\textbf{Efficient unrolling.} An imperative aspect to realize deep unfolding is the efficiency of unrolling with respect to the performance metrics. We use unrolling efficiency as an umbrella term encompassing performance indicators such as the convergence rate, computational complexity, inference time, and reliability. Several factors affecting the unrolling efficiency are the determination of trainable parameters, estimation and mapping of any non-linear functions of the iterative algorithms to equivalent activation functions, number of layers in the unrolled network, loss functions, and training process. Meanwhile, the need for efficiently unrolled intelligent modules will be pervasive across the 6G communication architecture. Hence, careful determination of the unrolling factors is an essential and vast research problem.

\textbf{Interoperability and Security} We expect O-RAN architectures to possess intelligent unfolded signal processing modules that can interface with the different O-RAN components to ensure stable and reliable operation. Further, maintaining openness and flexibility to support components from different vendors is essential along with interoperability with legacy systems. Additionally, when the solutions move towards open architecture and with deep learning depending heavily on data for training both offline and online, there will be an emergence of new security concerns. Consequently, there needs to be a deeper investigation into the security risks, new attack surfaces involved, and mitigation plans to keep up with the integration of deep learning based modules into core communication systems. 

\textbf{Hardware-efficient ML at the edge.} Edge learning will be a primary enabler in 6G networks. Devices that can perform self-optimization and act as intelligent decision agents without relying on a centralized cloud/fog server will be key to attain the undetectable latency and processing delay requirements envisioned for 6G communications. The power consumption of the devices serves as a key factor in attaining this capability. Hence, the intelligent architectures must be designed to be lightweight (hardware-efficient) involving few trainable parameters and layers while ensuring the desired reliability and computational performances.

We believe and hope that several of these challenges will be overcome and deep unfolded signal processing approaches will become one of the key enablers for 6G communication networks in the upcoming decade. 

\ifCLASSOPTIONcaptionsoff
  \newpage
\fi



%
\bibliographystyle{IEEEtran}	
\bibliography{Andro}

\begin{IEEEbiography}[{\includegraphics[width=1in,height=1.25in,clip,keepaspectratio]{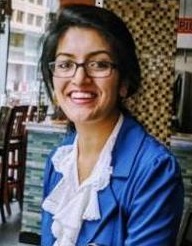}}]{Anu Jagannath} currently serves as the Founding Associate Director of Marconi-Rosenblatt AI/ML Innovation Lab at ANDRO Computational Solutions, LLC. She received her MS degree from State University of New York at Buffalo in Electrical Engineering. She is also a part-time PhD candidate with the Dept. of Electrical and Computer Engineering at Northeastern University, USA. Her research focuses on MIMO communications, Deep Machine Learning, Reinforcement Learning, Adaptive signal processing, Software Defined Radios, spectrum sensing, adaptive Physical layer, and cross layer techniques, medium access control and routing protocols, underwater wireless sensor networks, and signal intelligence. She has rendered her reviewing service for several leading IEEE conferences and Journals. She is the co-Principal Investigator (co-PI) and Technical Lead in multiple Rapid Innovation Fund (RIF) and SBIR/STTR efforts involving  applied AI/ML for wireless communications. She is also the inventor on 6 US Patents (granted,pending).
\end{IEEEbiography}

\begin{IEEEbiography}[{\includegraphics[width=1in,height=1.5in,keepaspectratio]{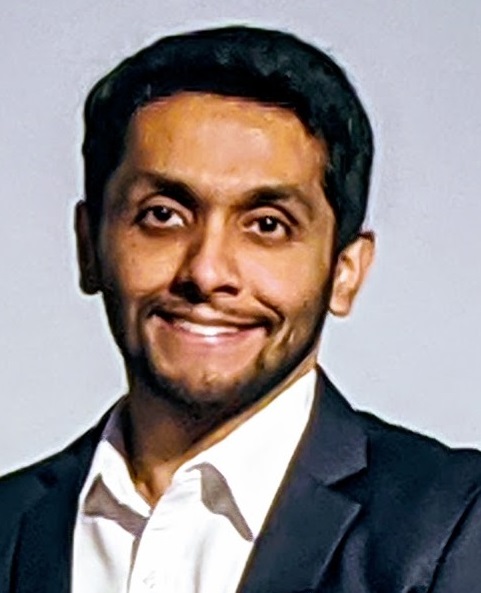}}]{Jithin Jagannath} (SM'19) is the Chief Technology Scientist and the Founding Director of the Marconi-Rosenblatt AI/ML Innovation Lab at ANDRO Computational Solutions, LLC. He is also the Adjunct Assistant Professor in the Department of Electrical Engineering at the University at Buffalo, State University of New York. He received his M.S. degree in Electrical Engineering from University at Buffalo; and received his Ph.D. degree in Electrical Engineering from Northeastern University. Dr. Jagannath heads several of the ANDRO's research and development projects in the field of signal processing, cognitive radio, cross-layer optimization, Internet-of-Things, and machine learning. He has been the lead author and Principal Investigator (PI) of several multi-million dollar research projects. He is currently leading several teams developing commercial products such as SPEARLink\texttrademark  among others. He is an IEEE Senior member and serves on IEEE Signal Processing Society's Applied Signal Processing Systems Technical Committee member. He is the inventor of 8 U.S. Patents (granted, pending). He also renders his service as Editor, TPC Member, and Reviewer to several leading journals and conferences.
\end{IEEEbiography}

\begin{IEEEbiography}[{\includegraphics[width=1in,height=1.25in,clip,keepaspectratio]{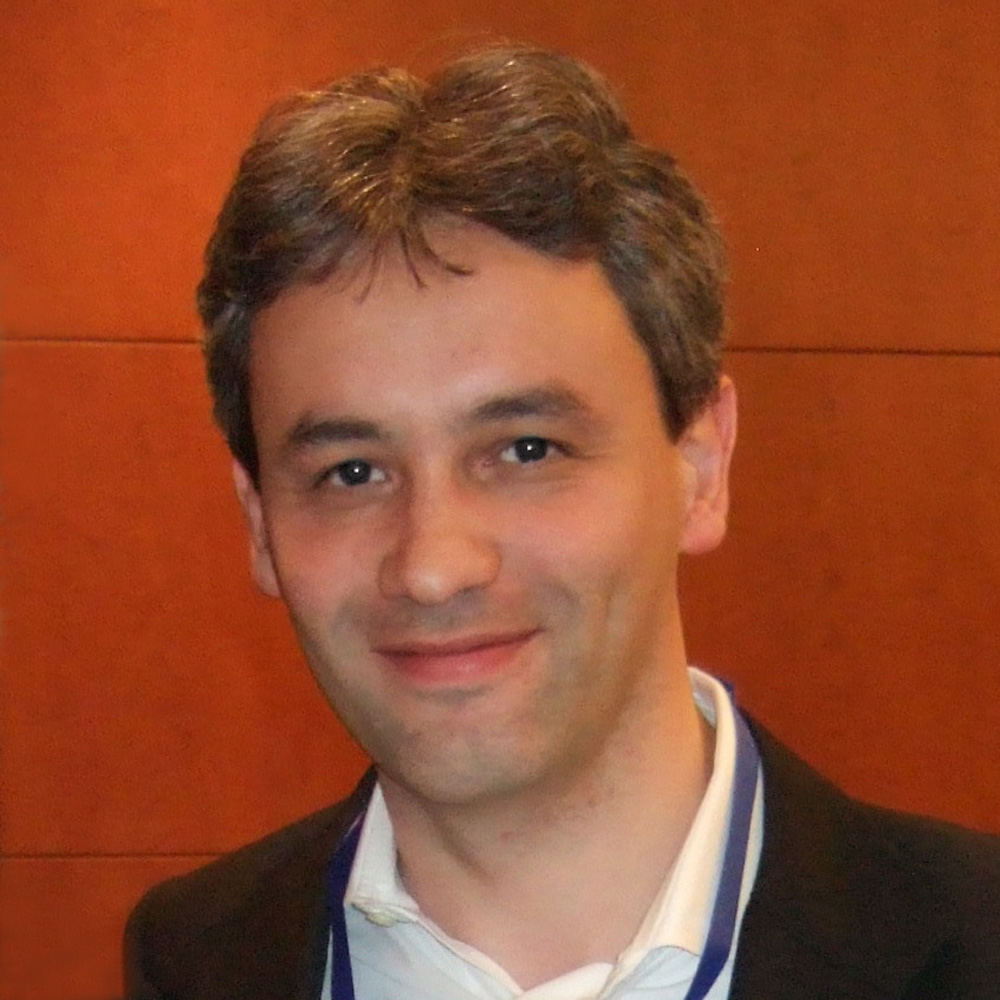}}]{Tomasso Melodia} (F'18) the William Lincoln Smith Professor with the Department of Electrical and Computer Engineering at Northeastern University in Boston. He is also the Founding Director of the Institute for the Wireless Internet of Things and the Director of Research for the PAWR Project Office. He received his Ph.D. in Electrical and Computer Engineering from the Georgia Institute of Technology in 2007. He is a recipient of the National Science Foundation CAREER award. Prof. Melodia is the Editor in Chief of Computer Networks. He has served as Technical Program Committee Chair for IEEE Infocom 2018, General Chair for IEEE SECON 2019, ACM Nanocom 2019, and ACM WUWnet 2014. Prof. Melodia is the Director of Research for the Platforms for Advanced Wireless Research (PAWR) Project Office, a \$100M public-private partnership to establish 4 city-scale platforms for wireless research to advance the US wireless ecosystem in years to come. Prof. Melodia's research on modeling, optimization, and experimental evaluation of Internet-of-Things and wireless networked systems has been funded by the National Science Foundation, the Air Force Research Laboratory the Office of Naval Research, DARPA, and the Army Research Laboratory. Prof. Melodia is a Fellow of the IEEE and a Senior Member of the ACM. 
\end{IEEEbiography}

\end{document}